\documentclass[%
 reprint,
 amsmath,amssymb,
 aps,
]{revtex4-1}

\usepackage{graphicx}
\usepackage{dcolumn}
\usepackage{bm}

\begin{document}


\title{Effect of color reconnection on forward-backward multiplicity and mean transverse momentum correlation}
\author{Sourav Kundu$^{1}$}
\email{souravkundu692@niser.ac.in}
\author{Bedangadas Mohanty$^{1,2}$}
\email{bedanga@niser.ac.in}
\author{Dukhishyam Mallick$^{1}$}
\email{dukhishyam.mallick@niser.ac.in}
\affiliation{$^{1}$School of Physical Sciences, National Institute of Science Education and Research, HBNI, Jatni 752050, India\\$^{2}$Experimental Physics Department, CERN - 1211, Geneva 23, Switzerland}
 
\date{\today}
\begin{abstract}

  Color reconnection (CR) mechanism in PYTHIA model has been reported to be essential to describe the flow-like collective effect observed in high multiplicity $p$+$p$ and $p$+Pb collisions. In this work, we test this mechanism towards explaining the Forward-Backward multiplicity correlation (b$_{\mathrm {cor}}$) measurements in $p$+$p$ collisions at the LHC energies. Out of the three different CR schemes implemented in PYTHIA, (a) MPI based CR (default mechanism), (b) QCD based CR and (c) Gluon moved CR, we found that the QCD based CR scheme describes relatively better the ALICE measurements of b$_{\mathrm {cor}}$ in $p$+$p$ collisions at $\sqrt{\mathrm s}$ = 0.9 and 7 TeV. In addition, we have tuned the parameters of the default CR mechanism in PYTHIA to describe simultaneously the measured charged particle multiplicity pseudo-rapidity ($\eta$) distribution and b$_{\mathrm {cor}}$. We found that an average number of multipartonic interactions ($\langle N_{\mathrm {MPI}} \rangle$)  between 2.5 to 3 and CR range between 0.9 to 2.5 best describes the experimental data.   Finally, we have presented a study using PYTHIA events for $p$+$p$ collisions at $\sqrt{\mathrm s}$ = 7 TeV which shows that the strength of Forward-Backward mean transverse momentum correlation (b$^{\langle p_{\mathrm{T}}\rangle \langle p_{\mathrm{T}} \rangle}_{\rm {cor}}$) is found to increase with CR in contrast to decrease of b$_{\rm {cor}}$ values with CR effect. Hence, simultaneously studying the b$^{\langle p_{\mathrm{T}}\rangle \langle p_{\mathrm{T}} \rangle}_{\rm {cor}}$ and b$_{\mathrm {\rm {cor}}}$ in the experiments will help in establishing the arguments either in favour or in disfavour of CR effect in the measurements.

\end{abstract}
\maketitle
\section{\label{sec:level1}Introduction}
In high-energy hadron or nuclei collisions, the study of correlations between produced particles are important to understand the dynamics of particle production mechanism. Several observables have been studied to understand the origin of the correlations between the produced particles. The Forward-Backward (F-B) correlations between charged particle multiplicity or mean transverse momentum ($ \langle p_{\mathrm{T}} \rangle$) in two separated pseudo-rapidity ($\eta$) windows are some of the observables.  F-B correlations allow us to decouple short range correlations (SRC) from long range correlations (LRC)~\cite{fb1,fb2}. SRC, which arises from short range effects such as decays of resonances or clusters, jet and mini-jet induced correlations, are typically localized effects extending in $\eta$ difference up to two units. Whereas the LRC that extends over a wider range of $\eta$ difference, originates from fluctuations associated with the particle emitting sources such as strings, clusters, cut pomerons, mini-jets~\cite{fb1,fb2,fb3,fb4}.

Forward-Backward multiplicity correlation (b$_{\rm {cor}}$) have been extensively studied by the UA5 Collaboration in $p$+$\bar{{p}}$ collisions at ISR energies from $\sqrt{\mathrm s}$ = 200 GeV to 900 GeV~\cite{data2,data22} and a linear correlation between multiplicity at forward and backward $\eta$ was reported. Later the correlation was confirmed by other collaborations in wider range of collisions energies 0.3 $<\sqrt{\mathrm s}<$ 1.8 TeV~\cite{data4,data44,data444,data5}. Recently ALICE~\cite{data6} and ATLAS~\cite{data7} collaboration at the LHC have reported detailed measurements of b$_{\rm {cor}}$ in $p$+$p$ collisions at $\sqrt{\mathrm s}$ = 0.9, 2.76 and 7 TeV.  A strong correlation between multiplicity at forward and backward $\eta$ has been reported. In contrast, measurements in e$^{+}$+e$^{-}$ collisions have reported a weak correlation~\cite{data55,data555}. This contrasting behaviour has attracted the attention of the theorists to understand the origin of b$_{\mathrm {cor}}$ in small system~\cite{sfm1,sfm2,fb5,fb6,fb7,fb8,fb9,qgsm}.

Further, observations in high multiplicity $p$+$p$ and $p$+Pb collisions at LHC energy show striking similarities between small system and heavy-ion collisions. New theoretical models have been proposed to understand these observations in high multiplicity $p$+$p$ and $p$+Pb collisions. It has been shown in Ref.~\cite{pythia_flow} that the collective flow-like behaviour in small systems, similar to those observed in heavy-ion collisions~\cite{alicebaryonmeson},  is explained by QCD based MC generator like PYTHIA~\cite{pythia1,pythia2}. It attributes the collectivity in the small systems to multipartonic interaction (MPI) and color reconnection (CR). CR in PYTHIA provides an alternate mechanism for flow-like effect~\cite{pythia_flow} in small systems compared to hydrodynamical processes attributed to flow in heavy-ion collisions~\cite{alicebaryonmeson,hydro1,hydro2,hydro3,hydro4,hydro5}. CR allows for the interaction between strings and this creates a flow-like effect in the final observable. In addition, the measured strong increase of $\langle p_{\mathrm{T}} \rangle$ of charged particles with event multiplicity is also attributed to the presence of CR mechanism between interacting color strings. Recently in Ref.~\cite{pythia_bcor}, it has been shown that the presence of CR is required for the qualitative explanation of the measured b$_{\mathrm {cor}}$.  Although a detailed study of b$_{\mathrm {cor}}$ in different CR models, which differ in the scheme by which string length is minimized, latest parametrization for CR and MPI is missing.

In this work, we have compared the values of b$_{\mathrm {cor}}$  calculated using different CR models implemented in PYTHIA with the corresponding experimental measurements. Default CR and MPI parametrizations in PYTHIA are further tuned with an attempt to explain the experimental measurements of b$_{\mathrm {cor}}$ and pseudo-rapidity ($\eta$) distribution simultaneously. Through this tuning process of the model parameters and using a $\chi^{2}$ minimization technique we have extracted the best fit values of  $\langle N_{\mathrm {MPI}} \rangle$ and CR range in $p$+$p$ collisions at $\sqrt{\mathrm s}$~=~7 TeV. We have also reported the first prediction for event shape dependence of b$_{\mathrm {cor}}$ using the PYHTIA model and the spherocity variable~\cite{spherocity}. Effect of CR on F-B correlation of intensive variable such as $\langle p_{\mathrm{T}} \rangle$ is also calculated. Simultaneous comparison of the measured b$_{\mathrm {cor}}$ and b$^{\langle p_{\mathrm{T}}\rangle \langle p_{\mathrm{T}} \rangle}_{\rm {cor}}$ with corresponding results from PYTHIA can shade light on the possible presence of CR as a physical phenomenon between the interacting strings and put more stringent constrains on the model parameters.

This work is organized as follows: In next section we have discussed about the PYTHIA model and the observables which are used for this study. In section III we present the results which include the comparison of b$_{\mathrm {cor}}$ between different CR models in PYTHIA and with experimental measurements, extraction of $\langle N_{\mathrm {MPI}} \rangle$ and CR range by simultaneous comparison of model to measurements of b$_{\mathrm {cor}}$  and $dN_{\mathrm {ch}}$/$d\eta$, prediction for the spherocity dependence of b$_{\mathrm {cor}}$, prediction for b$^{\langle p_{\mathrm{T}}\rangle \langle p_{\mathrm{T}} \rangle}_{\rm {cor}}$ in $p$+$p$ collisions at $\sqrt{\mathrm s}$ = 7 TeV. Finally, a summary of the work is presented in section IV.

\section{\label{sec:level2}Model description and F-B correlation observables}

\subsection{Model}
We have used PYTHIA event generator to simulate $p$+$p$ collision events at $\sqrt{\mathrm s}$ = 0.9 TeV and 7 TeV. PYTHIA is a monte carlo event generator and a combination of several models and theoretical calculations such as parton distributions, cross sections for hard and soft QCD process, multipartonic interactions,  color reconnection, initial and final-state parton showers, fragmentation and decay. For this study we have used PYTHIA 8.235, which is the latest tuned version of PYTHIA. This version includes MPI which is important for the explanation of multiplicity distributions~\cite{mpi1}, multiplicity dependent J/$\psi$ production~\cite{mpi2}, Jet and underlying event properties~\cite{mpi3} along with various other observables at LHC energies. This model also includes non perturbative effect like color reconnection which allows interaction between strings and creates a flow-like effect in the final observable. Details about the PYTHIA 8.235 version is given in Ref.~\cite{pythia1,pythia_ratio}.

We have generated 15 M inelastic events each for the 4 different schemes of CR implemented in PYTHIA. In one case CR effect is switched off and for other three cases CR effect is on. PYTHIA model has 3 different CR models. These models are MPI based CR model, QCD based CR model and Gluon move CR model. All these three CR models are built on the different strategy of minimization of the string length. MPI based CR model is the default CR model in PYTHIA, whereas QCD based CR model is the latest addition and termed here as CR new. One of the main difference between MPI based and QCD based CR model is the inclusion of the SU (3) colour rules from QCD during the reconnection process of strings and introduction of junction structures ~\cite{pythia1,pythia_ratio}. In gluon move CR model gluons can be moved from one location to another to reduce the total string length. In this model an additional optional mechanism (``flip'' mechanism) is also introduced to reduce string length. More details about the CR models can be found in~\cite{pythia1,pythia_ratio}.

\subsection{Observable}
F-B correlation is defined as follows:
\begin{equation}
b_{\mathrm {cor}} ~=~ \frac{\langle x_{\rm F}x_{\rm B}\rangle~-~\langle x_{\rm F}\rangle \langle x_{\rm B} \rangle}{\langle x^{2}_{\rm F} \rangle~-~\langle x_{\rm F} \rangle^{2}}
\end{equation}
Where $x$ is the observable. For b$_{\mathrm {cor}}$, $x$ is the multiplicity in an event in a particular acceptance and for b$^{\langle p_{\mathrm{T}}\rangle \langle p_{\mathrm{T}} \rangle}_{\rm {cor}}$ $x$ is the $\langle p_{\mathrm{T}} \rangle$ in an event for a given experimental acceptance. F corresponds to the forward $\eta$ window and B corresponds to the backward $\eta$ window. For this analysis following terminology is used:
\begin{itemize}
\item $\delta \eta$ = width of the $\eta$ window.
\item $\eta_{gap}$ = center to center distance between forward and backward $\eta$ window.
\item $\eta_{sep}$ = distance between the lower and upper boundary of forward and backward $\eta$ window.
\item $\phi_{sep}$ = center to center distance between forward and backward $\phi$ window.
\end{itemize}
This terminology is same as used in ALICE data analysis and details about it can be found in~\cite{data6}. In order to compare with the ALICE measurements, same kinematic selections as in data are used.

\section{\label{sec:level3}Results}

\subsubsection{\label{sec:level3}b$_{\rm {cor}}$ and color reconnection model}
\begin{figure*}[hbtp]
\centering 
\includegraphics[scale=0.7]{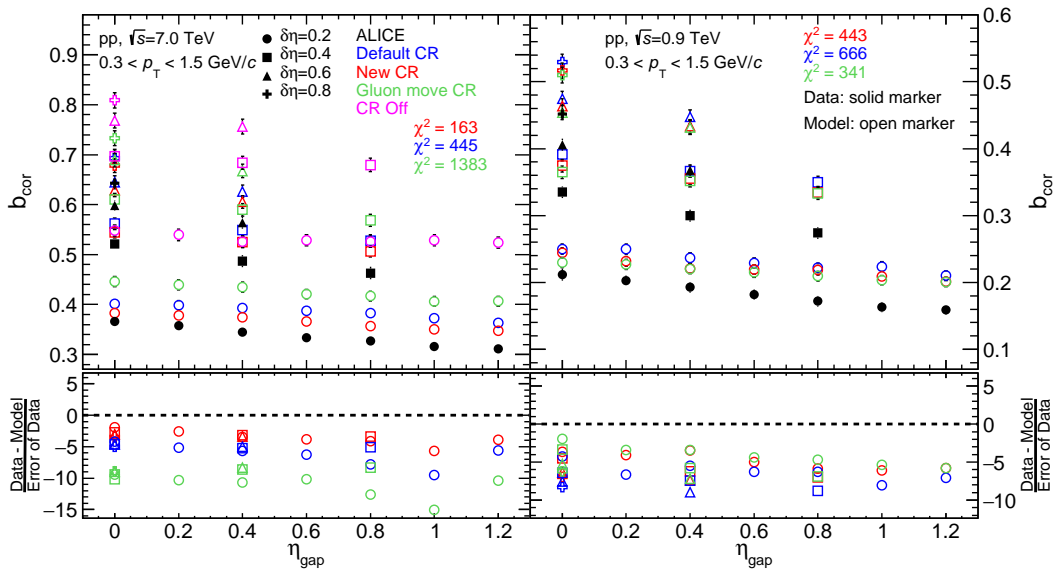}
\caption{(Color online) Left panel : b$_{\mathrm {cor}}$ vs. $\eta_{\mathrm {gap}}$ for different $\delta \eta$, in p+p collisions at $\sqrt{s}$ = 7 TeV for different CR models in PYTHIA. The model results are compared with the corresponding measurements from the ALICE experiment~\cite{data6}. Errors on the experimental data points are the quadrature sum of statistical and systematic errors. Lower panel of the plot shows the standard deviation between the data and model for the different CR schemes. Right panel: same as left panel but for $p$+$p$ collisions at $\sqrt{\mathrm s}$ = 0.9 TeV.}
\label{fig:fb1} 
\end{figure*}

Figure~\ref{fig:fb1} shows the b$_{\mathrm {cor}}$ as a function of $\eta_{\mathrm {gap}}$ for different $\delta \eta$, in $p$+$p$ collisions at $\sqrt{\mathrm s}$ = 7 TeV (left panel) and 0.9 TeV (right panel). Results for four different CR schemes: 1) No CR, 2) Default CR model, 3) new CR model and 4) Gluon move CR model in PYTHIA framework are also shown. 
Measurements from models are also compared with the experimental data.

We find that the CR is required to describe the experimental data on b$_{\mathrm {cor}}$. ALICE measurements show that the strength of b$_{\mathrm {cor}}$  increases with increasing collision energy, $\delta \eta$ and decreases with increasing $\eta_{\mathrm {gap}}$. All CR models are able to describe these trends of the data but they overestimate the values of the experimental measurements. However, new CR model is quantitatively more closer to the experimental measurements. The Gluon move CR model gives higher values of b$_{\rm {cor}}$ compared to other models in $p$+$p$ collisions at $\sqrt{s}$ = 7 TeV. Whereas at the low energy it gives similar values as the new CR model. PYTHIA model calculations indicate that the inclusion of the colour rules from QCD and introduction of junction structures~\cite{pythia1,pythia_ratio} in CR model provide a better description of the measured b$_{\mathrm {cor}}$ values as well as various particle ratios as reported in Ref.~\cite{pythia_ratio}. Comparison of data with PYTHIA model implies further tuning of the model parameters is required, in order for a more quantitative description of data. We have carried out one such tuning study using the default PYTHIA model.

In Ref.~\cite{pythia_bcor}, the authors have shown only using the default PYTHIA model with CR strength 0.9 together with different ranges of number of MPI at different collision energies can explain the ALICE measurements. However, the experimental data corresponds to minimum bias $p$+$p$ collisions, hence includes contributions from a distribution of MPI, so sharp cuts on number of MPI may not be a satisfactory approach. In addition, b$_{\mathrm {cor}}$ over an extensive quantity could strongly depend on the range of the event selection variables~\cite{bcor_pbpb}. One can get similar strength of b$_{\rm {cor}}$ by selecting two different ranges of number of MPI, and those may correspond to different values of average number of MPI for the entire event class.

\subsubsection{\label{sec:level3}Extraction of color reconnection range and average number of MPI}
In order to describe the ALICE measurements of b$_{\mathrm {cor}}$ we have used the default CR model which has parameters corresponding to the number of MPI and CR range. We have varied the range of CR in default PYTHIA model and for each of the variations we have generated different MPI distribution by varying the parameter ''MultipartonInteractions:pT0Ref''. Probability distribution of MPI depends on the 2$\rightarrow$2 parton-parton cross-section, which diverges for p$_{T}~\rightarrow~$0. To avoid this divergence the parton-parton cross-section is modelled as,\\
\begin{equation}
\frac{d\sigma}{dp^{2}_{T}} ~\propto~ \frac{\alpha_{s}(p^{2}_{T})}{p^{4}_{T}}~\rightarrow~ \frac{\alpha_{s}(p^{2}_{T}+p^{2}_{T0})}{(p^{2}_{T}+p^{2}_{T0})^{2}}.
\end{equation}
Here p$_{T0}$ is a model parameter in PYTHIA that governs the MPI distribution. p$_{T0}$ is further modelled as an energy dependent parameter,\\
\begin{equation}
p_{T0}=p^{\rm {ref}}_{T0}(\frac{E_{\rm {cm}}}{E^{\rm {ref}}_{\rm {cm}}})^{n}.
\end{equation}
Here p$^{\rm {ref}}_{\rm {T0}}$ (default = 2.28 GeV/$c$), E$^{ref}$$_{\rm {cm}}$ (default = 7000 GeV) and $n$ (default = 0.215) are input parameters which control p$_{T0}$, hence the MPI distribution and average number of MPI ($\langle N_{\mathrm {MPI}} \rangle$). Parameter p$_{\rm {T0}}$ is also appears in the probability distribution of CR, which is defined as,\\
\begin{equation}
P(CR)=\frac{R*p_{T0}}{R^{2}*p^{2}_{\rm {T0}}+p^{2}_{\rm T}}.
\end{equation}
Where R is the CR range.

\begin{figure}[hbtp]
\centering 
\includegraphics[scale=0.4]{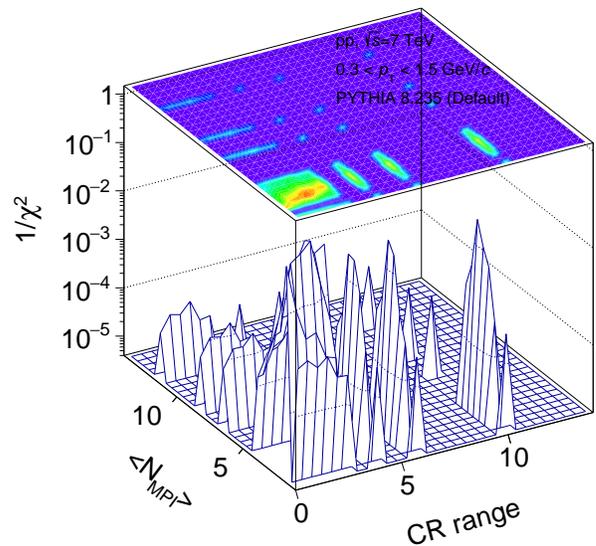}
\caption{(Color online) 1/$\chi^{2}$ as a function of $\langle N_{\rm {MPI}} \rangle$ and CR range in p+p collisions at $\sqrt{\rm s}$ = 7 TeV. $\chi^{2}$ values are estimated by comparing the PYTHIA results with ALICE measurements of $\eta$ distribution~\cite{dataeta} and b$_{\rm {cor}}$~\cite{data6} for different $\eta_{\rm {gap}}$ and $\delta \eta$.}.
\label{fig:xi} 
\end{figure}

The $p$+$p$ collision events at $\sqrt{s}$ = 7 TeV have been generated by varying both CR range and p$^{\rm {ref}}_{\rm {T0}}$. Then the charged particle multiplicity $\eta$ distribution and b$_{\mathrm {cor}}$ values for different $\eta_{\mathrm {gap}}$ and $\delta \eta$ are computed.  The results for each event set (with a fixed value of p$^{\rm {ref}}_{\rm {T0}}$ and CR range) from PYTHIA model is compared with the corresponding ALICE measurements and a combined $\chi^{2}$ value is calculated. Figure~\ref{fig:xi} shows 1/$\chi^{2}$ as a function of different combinations of p$^{\rm {ref}}_{\rm {T0}}$ and CR range used for the PYTHIA events. The minimum $\chi^{2}$ occurs for p$^{\rm {ref}}_{\rm {T0}}$ = 2.6 GeV/$c$ and CR range = 1.5. However there are few other combinations of p$^{\rm {ref}}_{\rm {T0}}$ and CR range which also give the $\chi^{2}$/ndf~$\le$~1 for the fitting of both b$_{\mathrm {cor}}$ and $\eta$ distributions and these are listed in Table~\ref{table:comp}. From our study we can conclude that in order to simultaneously describe both $\eta$ distribution and b$_{\rm {cor}}$ values for charged particles in p+p collisions at $\sqrt{s}$ = 7 TeV, we need $\langle N_{\mathrm {MPI}} \rangle$ values between 2.5 to 3.0 and CR range between 0.9 to 2.5. The parametrization listed as Tune 1 is used for further comparison with default parametrization in the rest of the paper.

\begin{table}[hbtp]
\caption{The p$^{\rm {ref}}_{\rm {T0}}$  parameter, $\langle N_{\rm {MPI}} \rangle$ and CR range in PYTHIA that best describes the measured charged particle $\eta$ distribution and b$_{\mathrm {cor}}$  in p+p collisions at $\sqrt{s}$ = 7 TeV.}
\label{table:comp}
\begin{center}
\begin{tabular}{|c|c|c|c|}
\hline
Case&p$^{ref}_{T0}$ (GeV/$c$)&$<N_{MPI}>$&CR range\\
\hline
Default&2.28&3.5&1.8\\
\hline
Tune1&2.6&2.7&1.5\\
\hline
Tune2&2.6&2.7&1.6\\
\hline
Tune3&2.7&2.5&0.9\\
\hline
Tune4&2.5&2.9&2.5\\
\hline
\end{tabular}
\end{center}
\end{table}

\begin{figure}[hbtp]
\centering 
\includegraphics[scale=0.4]{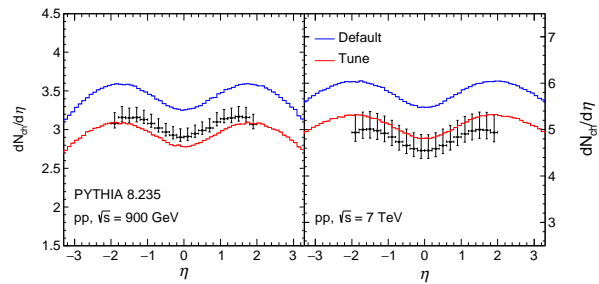}
\caption{(Color online) The charged particle $\eta$ distribution measured in ALICE~\cite{dataeta}
 for p+p collisions at $\sqrt{\rm s}$ = 7 TeV and 0.9 TeV are compared to the PYTHIA model calculations with default and tuned model parameters.}
\label{fig:eta} 
\end{figure}

Figure~\ref{fig:eta} shows comparison of measured $\eta$ distribution of charged particles in ALICE with the results from PYTHIA model calculation $p$+$p$ collisions at $\sqrt{\rm s}$ = 7 TeV (right panel) and 0.9 TeV (left panel).  The default and tune corresponds to the parameter values of p$^{\rm {ref}}_{\rm {T0}}$ and CR range given in Table~\ref{table:comp}. The PYTHIA tuned shows a better agreement with the data.

\begin{figure*}[hbtp]
\centering 
\includegraphics[scale=0.7]{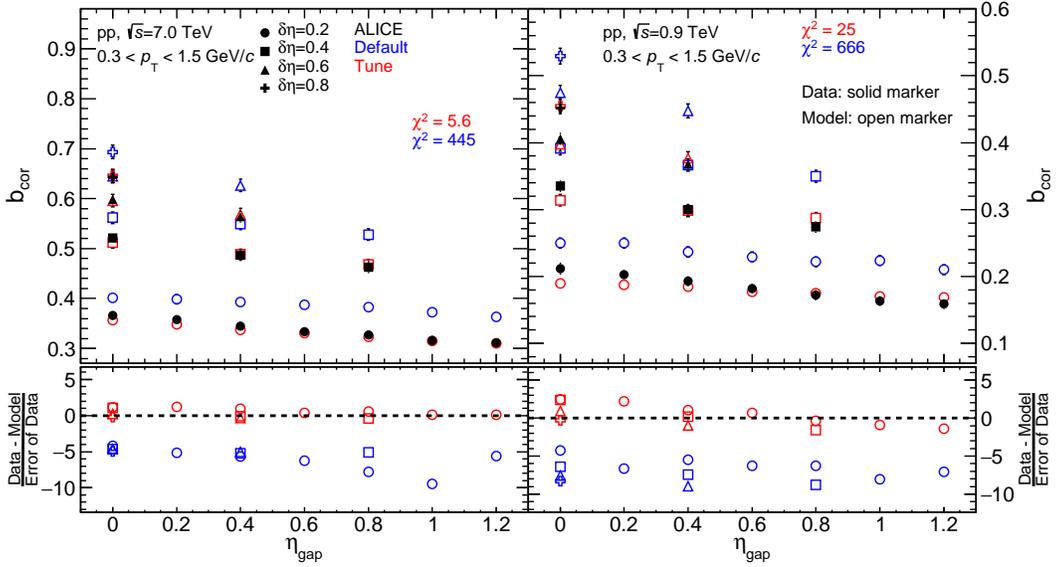}
\caption{(Color online) Left panel : The ALICE measurements of b$_{\rm {cor}}$ vs. $\eta_{\rm {gap}}$ for different $\delta \eta$, in $p$+$p$ collisions at $\sqrt{\rm s}$ = 7 TeV~\cite{data6} compared to both default and tune parametrization in PYTHIA. Errors on the experimental data points are the quadrature sum of statistical and systematic errors. Lower panel of the plot shows the standard deviation for both the cases. Right panel: same as left panel but for p+p collisions at $\sqrt{\rm s}$ = 0.9 TeV.}
\label{fig:fb2} 
\end{figure*}

Figure~\ref{fig:fb2} shows the b$_{\mathrm {cor}}$ values as a function of $\eta_{\mathrm {gap}}$ for different $\delta \eta$, in $p$+$p$ collisions at $\sqrt{\rm s}$ = 7 TeV (left panel) and 0.9 TeV (right panel) from ALICE and PYTHIA. Results from the ALICE measurements are compared to both default and tune case in PYTHIA. The PYTHIA tuned describes the measurements better compared to default case. Further, the PYTHIA model parameters p$^{\rm {ref}}_{\rm {T0}}$  and CR range which were tuned by comparing to the measured charged particle $\eta$ distribution and b$_{\mathrm {cor}}$ measurements in p+p collisions at $\sqrt{\rm s}$ = 7 TeV are found to provide a good description of the results in p+p collisions at $\sqrt{\rm s}$ = 0.9 TeV compared to default parametrization.

\begin{figure*}[hbtp]
\centering 
\includegraphics[scale=0.7]{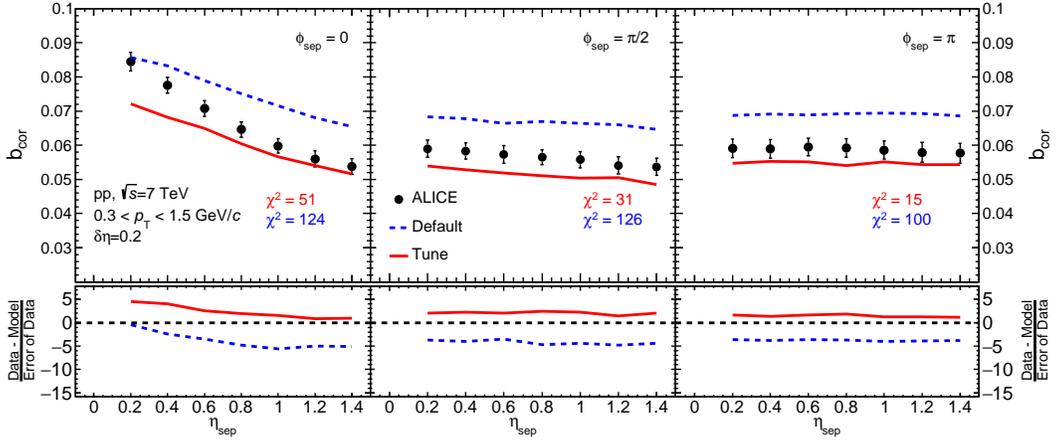}
\caption{(Color online) The ALICE measurements of b$_{\rm {cor}}$ vs. $\eta_{\rm {sep}}$ for $\delta \eta$ = 0.2 with 3 different $\phi_{\rm {sep}}$, in $p$+$p$ collisions at $\sqrt{\rm s}$ = 7 TeV~\cite{data6} compared to different parametrization of model parameters in PYTHIA. Errors on the experimental data points are the quadrature sum of statistical and systematic errors. Lower panel of the plot shows the standard deviation between the data and the model for the different tunes.}
\label{fig:fb3} 
\end{figure*}

ALICE has also measured the F-B multiplicity correlations in different azimuthal sectors. These azimuthal sectors in forward and backward $\eta$ have $\delta \eta$ = 0.2 and $\delta \phi$ = $\pi /4$. There are 5 different combinations of $\phi_{\rm {sep}}$. More details about the selection of analysis window can be found in Ref.~\cite{data6}. Figure~\ref{fig:fb3} shows the ALICE measurements of b$_{\rm {cor}}$ as a function of $\eta_{\rm {sep}}$ for $\delta \eta$ = 0.2 and 3 different $\phi_{\rm {sep}}$ in p+p collisions at $\sqrt{\rm s}$ = 7 TeV. Experimental measurements are compared with the both PYTHIA default and tuned. PYTHIA tuned compares better with data than the default case.

\subsubsection{\label{sec:level3}b$_{\rm {cor}}$ and spherocity}
\begin{figure*}[hbtp]
\centering 
\includegraphics[scale=0.6]{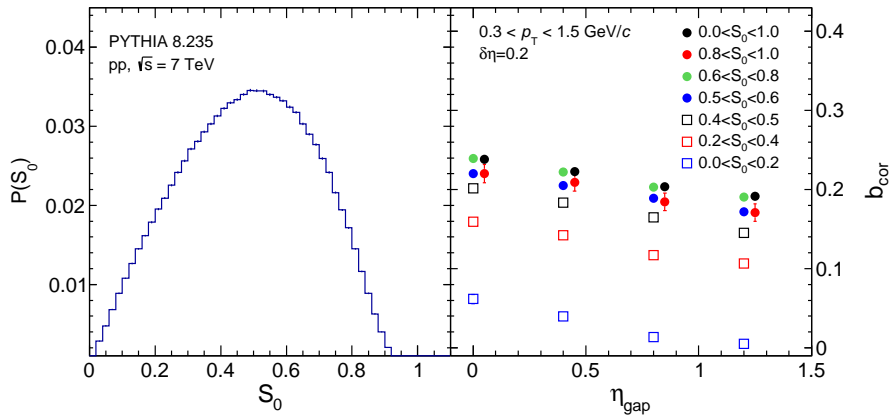}
\caption{(Color online) Left panel : Spherocity distribution in $p$+$p$ collisions at $\sqrt{\rm s}$ = 7 TeV from PYTHIA model. Right panel : b$_{\rm {cor}}$ vs. $\eta_{\rm {gap}}$ for $\delta \eta$ = 0.2 for different spherocity intervals  in $p$+$p$ collisions at $\sqrt{\rm s}$~=~7~TeV from PYTHIA.}
\label{fig:fb4} 
\end{figure*}

A spherocity dependent study of b$_{\rm {cor}}$ is performed to understand the source of multiplicity correlation in p+p collisions. Spherocity variable is defined~\cite{spherocity} as follows\\
\begin{equation}
S_{0} ~=~\frac{\pi^{2}}{4} \Bigg(\frac{\sum_{i} \vec{p}_{\mathrm{T_{i}}~\times~\hat{n}}}{\sum_{i} \vec{p}_{\mathrm{T_{i}}}}\Bigg)^{2},
\end{equation}
where $p_{\mathrm{T}}$ is the transverse momentum and $\hat{n}$ is a unit vector which minimizes $S_{0}$. For spherocity dependent analysis we have chosen those events which have at least 3 charged particles at $|\eta|~<$ 0.8. Spherocity variable allow us to distinguish isotropic or spherical events where MPI is dominant, from the pencil shaped jetty events which are dominated with dijets. Low value of $S_{0}$ corresponds to the jetty events and high value of $S_{0}$ corresponds to the spherical events~\cite{spherocity}. If $S_{0}$~=~0, then the event is completely jetty and if $S_{0}$~=~1, then the event is fully isotropic. Left panel of Fig.~\ref{fig:fb4} shows the spherocity distribution for charged particles in $p$+$p$ collisions at $\sqrt{\rm s}$~=~7~TeV from PYTHIA. The calculated b$_{\rm {cor}}$ for different $S_{0}$ intervals are shown in the right panel of Fig.~\ref{fig:fb4} as a function of $\eta_{\rm {gap}}$ for $\delta \eta$ = 0.2. The value of b$_{\rm {cor}}$ (or strength of F-B correlations) is higher for isotropic events compared to jetty events. As MPI is dominant for isotropic events, it seems most of the contribution to the value of b$_{\rm {cor}}$ can be attributed to MPI. Such a calculation can be easily performed in experiments and it will throw light on MPI as a source of F-B correlations.

\subsubsection{\label{sec:level3}b$^{\langle p_{\mathrm{T}}\rangle \langle p_{\mathrm{T}} \rangle}_{\rm {cor}}$ and color reconnection}
In this section we have studied the effect of CR on the F-B correlation of an intensive variable. For intensive variable we have chosen $\langle p_{\mathrm{T}} \rangle$. We have seen from the previous CR model studies that the $p_{\mathrm {T}}$ spectra get harder in presence of CR, which causes an enhancement of $\langle p_{\mathrm{T}} \rangle$~\cite{pythia_flow}. This enhancement in the $\langle p_{\mathrm{T}}\rangle$ for small systems mainly has been attributed to collective flow-like due to the presence of CR~\cite{pythia_flow} . Therefore we also expect effect of CR on F-B $\langle p_{\mathrm{T}} \rangle$ correlation (b$^{\langle p_{\mathrm{T}}\rangle \langle p_{\mathrm{T}} \rangle}_{\rm {cor}}$).

\begin{figure*}[hbtp]
\centering 
\includegraphics[scale=0.6]{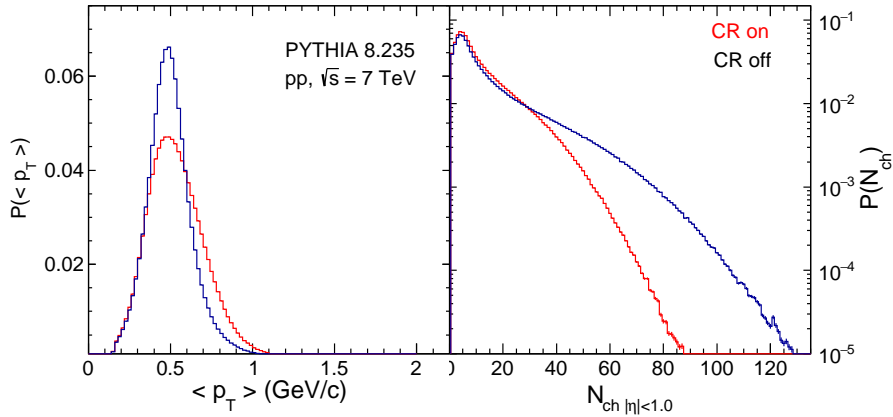}
\caption{(Color online) Left panel : Event-by-event $\langle p_{\mathrm{T}}\rangle$ distribution in $p$+$p$ collisions at $\sqrt{\rm s}$ = 7 TeV from PYTHIA model with CR on and CR off. Right panel : Same as left panel but for event-by-event charged particle multiplicity distribution.}
\label{fig:fb5} 
\end{figure*}
Left panel of Fig.~\ref{fig:fb5} shows the event-by-event $\langle p_{\mathrm{T}} \rangle$ distribution and the right panel of Fig.~\ref{fig:fb5} shows the event-by-event multiplicity (N$_{\rm {ch}}$) distribution from PYTHIA for two different cases: CR on and CR off. Interestingly we find that the CR has opposite effects on $\langle p_{\mathrm{T}} \rangle$ and N$_{\rm {ch}}$. In presence of CR event-by-event $\langle p_{\mathrm{T}} \rangle$ distribution gets wider whereas the event-by-event N$_{\rm {ch}}$ distribution gets narrower. This observation indicates that the effect of CR on b$^{\langle p_{\mathrm{T}}\rangle \langle p_{\mathrm{T}} \rangle}_{\rm {cor}}$ and b$_{\rm {cor}}$ will probably also be opposite.

\begin{figure}[hbtp]
\centering 
\includegraphics[scale=0.4]{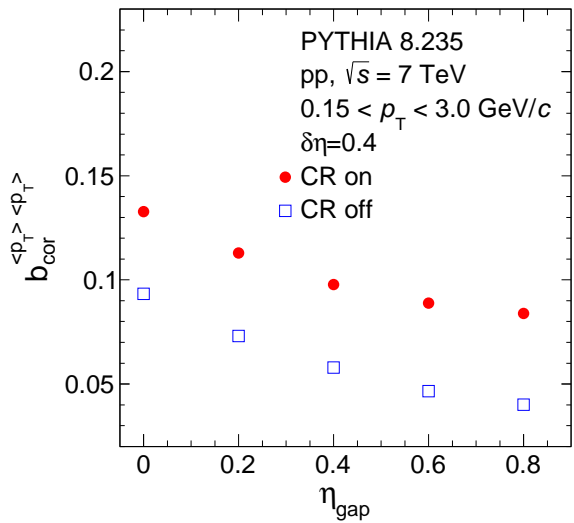}
\caption{(Color online)  b$^{\langle p_{\mathrm{T}}\rangle \langle p_{\mathrm{T}} \rangle}_{\rm {cor}}$ vs. $\eta_{\rm {gap}}$ for $\delta \eta$ = 0.4 in p+p collisions at $\sqrt{\rm s}$ = 7 TeV from PYTHIA model study for CR on and CR off configuration.}
\label{fig:fb6} 
\end{figure}
Figure~\ref{fig:fb6} shows b$^{\langle p_{\mathrm{T}}\rangle \langle p_{\mathrm{T}} \rangle}_{\rm {cor}}$ as a function of $\eta_ {\rm{gap}}$ for $\delta \eta$~=~0.4 in $p$+$p$ collisions at $\sqrt{\rm s}$ = 7 TeV from PYTHIA model study for CR on and CR off. As observed for b$_{\rm {cor}}$ versus $\eta_ {\rm{gap}}$, the  b$^{\langle p_{\mathrm{T}}\rangle \langle p_{\mathrm{T}} \rangle}_{\rm {cor}}$ value also decreases with increasing $\eta_{\rm {gap}}$. However, the presence of CR gives larger values of b$^{\langle p_{\mathrm{T}}\rangle \langle p_{\mathrm{T}} \rangle}_{\rm {cor}}$ (increased F-B correlations) in striking contrast to the  effect of CR on b$_{\rm {cor}}$ which decreases the values (reduces F-B correlations). Therefore, simultaneous measurement of b$^{\langle p_{\mathrm{T}}\rangle \langle p_{\mathrm{T}} \rangle}_{\rm {cor}}$ and b$_{{\rm cor}}$ in  experiments and comparison with CR model will strengthen the arguments for the possible presence of CR effect between the interacting strings as implemented in PYTHIA. 

\section{\label{sec:level4}Summary}
We have presented the multiplicity and $\langle p_{\mathrm{T}} \rangle$ correlation of the produced charged particles in forward and backward pseudo-rapidity windows, using PYTHIA event generator for $p$+$p$ collisions at $\sqrt{s}$ = 0.9 and 7 TeV. ALICE measurements for b$_{\rm {cor}}$ is compared with three different CR models as implemented in PYTHIA. A tuning of model parameters related to $\langle N_{\rm{MPI}} \rangle$ and CR range in default PYTHIA model is done for the quantitative description of the data. Through this exercise the  $\langle N_{\rm{MPI}} \rangle$ and CR range that best describes the LHC data are estimated   We have also performed a spherocity dependent analysis to separate the contribution to b$_{\rm {cor}}$ from the isotropic events and jetty events. We find that the F-B correlations in the PYTHIA model for isotropic events are dominantly due to MPI effects. F-B correlation for $ \langle p_{\mathrm{T}} \rangle$ is also presented and effect of CR on F-B $\langle p_{\mathrm{T}} \rangle$ correlation is estimated. Main conclusions from our study are:\\
\begin{itemize}
\item In PYTHIA model framework, CR mechanism is needed to describe the ALICE measurements of F-B correlations. All models of CR as implemented in PYTHIA over estimates the experimentally measured b$_{\rm {cor}}$ values. However, the new CR model which include colour rules from QCD as an additional constraint during the reconnection of the strings and  junction structures~\cite{pythia1,pythia_ratio} qualitatively describe the data better compared to other methods of implementation of CR effect in PYTHIA. 
\item The experimental measured charged particle $\eta$ distribution and F-B correlations for $p$+$p$ collisions at $\sqrt{\rm s}$ = 7 TeV was used simultaneously to tune and get the best fit model parameters for the default CR model in PYTHIA. A $\chi^{2}$ minimization method yield the best fit model parameters corresponding to $\langle N_{\rm {MPI}} \rangle$ values between 2.5 to 3.0 and CR range between 0.9 to 2.5. These values were also found to reasonably describe the measurements of $dN_{\rm{ch}}/d\eta$ distribution and b$_{\rm {cor}}$ values for $p$+$p$ collisions at $\sqrt{\rm s}$ = 0.9 TeV.
\item A spherocity based analysis of PYTHIA events showed that the F-B correlations are dominantly coming from isotropic events and their source are MPI effects. It will be interesting to carry out such measurements in experiments, as it can provide more inputs to contrain the model parameters.
\item One of the most interesting findings is the observation of CR having opposite effect on b$^{\langle p_{\mathrm{T}}\rangle \langle p_{\mathrm{T}} \rangle}_{\rm {cor}}$ compared to b$_{\rm {cor}}$ in PYTHIA. In presence of CR effects $\langle p_{\mathrm{T}} \rangle$ distribution is narrower and  b$^{\langle p_{\mathrm{T}}\rangle \langle p_{\mathrm{T}} \rangle}_{\rm {cor}}$ values larger compared to CR off case. This is in sharp contract to $N_{\rm{ch}}$ distributions being wider and b$_{\rm {cor}}$ values smaller for CR on case relative to CR off case. We encourage simultaneous experimental measurements of b$^{\langle p_{\mathrm{T}}\rangle \langle p_{\mathrm{T}} \rangle}_{\rm {cor}}$ and b$_{\rm {cor}}$,  as it will not only help to constrain model parameters but also possibly validate the CR mechanism between the interacting strings as implemented in PYTHIA. 
\end{itemize}

In future, from the model perspective, we would like to carry out a study of  b$^{\langle p_{\mathrm{T}}\rangle \langle p_{\mathrm{T}} \rangle}_{\rm {cor}}$ as a function of charged particle multiplicity, as this variable is robust under multiplicity and centrality selection.  Further, we would like to study b$^{\langle p_{\mathrm{T}}\rangle \langle p_{\mathrm{T}} \rangle}_{\rm {cor}}$ for identified particles as it would be interesting in connection to CR effects explaining the flow-like behaviour seen in high-energy and high multiplicity $p$+$p$ collisions.

\section{\label{sec:level5}Acknowledgements}
B.M. acknowledges the financial support from J C Bose National Fellowship of DST, Government of India.


\begin{thebibliography}{50}
\bibitem{fb1} A. Capella, A. Krzywicki, Phys. Rev. D 18, 4120 (1978).
\bibitem{fb2} A. Capella, U. Sukhatme, C. I. Tan, J. T. T. Van, Phys. Rep. 236, 225 (1994).
\bibitem{fb3} M. Braun, R. Kolevatov, C. Pajares, V. Vechernin, Eur. Phys. J. C 32, 535 (2004).
\bibitem{fb4} N. Amelin, N. Armesto, M. Braun, E. Ferreiro, C. Pajares, Phys. Rev. Lett. 73, 2813 (1994).
\bibitem{data1} British-French-Scandinavian Collaboration, M. Albrow et al., Nucl. Phys. B 145, 305 (1978).
\bibitem{data2} UA5 Collaboration, K. Alpgard et al., Phys. Lett. B 123, 361 (1983).
\bibitem{data22} UA5 Collaboration, G. G. Alner et al., Phys. Rep. 154, 247 (1987).
\bibitem{data4} E735 Collaboration, T. Alexopoulos et al., Phys. Lett. B 353, 155 (1995).
\bibitem{data44} L. V. Bravina et al., Sov. J. Nucl. Phys. 50, 245 (1989).
\bibitem{data444} V. V. Aivazyan et al., NA22 Collaboration, Z. Phys. C 42, 533 (1989).
\bibitem{data5} STAR Collaboration, B. Abelev et al., Phys. Rev. Lett. 103, 172301 (2009).
\bibitem{data6} ALICE Collaboration, J. Adam et al., JHEP 1505, 097 (2015).
\bibitem{data7} ATLAS Collaboration, G. Aad et al., JHEP 1207, 019 (2012).
\bibitem{data55} W. Braunschweig et al., TASSO Collaboration, Z. Phys. C 45, 193 (1989).
\bibitem{data555} R. Akers et al., OPAL Collaboration, Phys. Lett. B 320, 417 (1994).
\bibitem{sfm1} M. Braun and C. Pajares, Phys. Lett. B 287, 154 (1992).
\bibitem{sfm2} M. Braun and C. Pajares, Nucl. Phys. B 390, 542 (1993).
\bibitem{fb5} A. Dumitru, F. Gelis, L. McLerran, R. Venugopalan, Nucl. Phys. A 810, 91 (2008).
\bibitem{fb6} T. S. Biro, H. B. Nielsen, J. Knoll, Nucl. Phys. B 245, 449 (1984).
\bibitem{fb7} A. Bialas, W. Czyz, Nucl. Phys. B 267, 242 (1986).
\bibitem{fb8} M. A. Braun, C. Pajares, Phys. Lett. B 287, 154 (1992).
\bibitem{fb9} M. A. Braun, C. Pajares, Nucl. Phys. B 390, 542 (1993).
\bibitem{qgsm} L. V. Bravina, J. Bleibel, E. E. Zabrodin Phys. Lett. B 787, 146 (2018).
\bibitem{pythia_flow} A. Ortiz, P. Christiansen, E. Cuautle, I. Maldonado, G. Paic, Phys. Rev. Lett. 111, 042001 (2013).
\bibitem{alicebaryonmeson} ALICE Collaboration, B. Abelev et al., Phys. Rev. Lett. 111, 222301 (2013).
\bibitem{pythia1} T. Sj{\"o}strand, S. Ask, J. R. Christiansen, R. Corke, N. Desai, P. Ilten, S. Mrenna, S. Prestel, C. O. Rasmussen, P. Z. Skands, Comput. Phys. Commun. 191, 159 (2015).
\bibitem{pythia2} P. Skands, S. Carranza, J. Rojo, Eur. Phys. J. C 74, 3024 (2014).
\bibitem{hydro1} J. D. Bjorken, Phys. Rev. D 27, 140 (1983).
\bibitem{hydro2} C. M. Hung and E. V. Shuryak, Phys. Rev. C 57, 1891 (1998).
\bibitem{hydro3} S. A. Voloshin and A. M. Poskanzer, Phys. Lett. B 474, 27 (2000).
\bibitem{hydro4} H. Song and U. Heinz, Phys. Rev. C 77, 064901 (2008).
\bibitem{hydro5} E. Cuautle and G. Paic, J. Phys. G 35, 075103 (2008).
\bibitem{pythia_bcor} E. Cuautle, E. Dominguez, I. Maldonado, Eur. Phys. J. C 79, 626 (2019).
\bibitem{spherocity} A. Banfi, G. P. Salam, and G. Zanderighi, JHEP 06, 038 (2010).
\bibitem{mpi1} T. Sj{\"o}strand and M. V. Zijl, Phys. Rev. D 36, 2019 (1987).
\bibitem{mpi2} S. G. Weber, A. Dubla, A. Andronic and A. Morsch, Eur. Phys. J. C 79, 36 (2019). 
\bibitem{mpi3} CMS Collaboration, S. Chatrchyan et al., Eur. Phys. J. C 73, 2674 (2013).
\bibitem{pythia_ratio} C. Bierlich and J. R. Christiansen, Phys. Rev. D 92, 094010 (2015).
\bibitem{bcor_pbpb} I. Altsybeev (for the ALICE Collaboration), KnE Energy and Physics, 304 (2018).
\bibitem{dataeta} ALICE Collaboration, J. Adam, et al.,  Eur. Phys. J. C 77, 33 (2017).
\end{thebibliography}
\end{document}